\documentclass[pre,aps,floatfix, superscriptaddress,showpacs,twocolumn]{revtex4-1}
\usepackage{amsmath,bm,graphicx}
\usepackage{amssymb}
\usepackage{amsfonts}
\usepackage{natbib}
\usepackage[dvipsnames]{color}
\usepackage{ulem}

\newcommand{\be}{\begin{equation}} 
\newcommand{\ee}{\end{equation}}
\newcommand{\bea}{\begin{eqnarray}}   
\newcommand{\eea}{\end{eqnarray}}

\newcommand{\rr}{{\bf r}}
\newcommand{\NN}{{\bf \nabla}}
\newcommand{\FF}{{\bf F}}

\newcommand{\na}{n^{\alpha}}
\newcommand{\nb}{n^{\beta}}

\newcommand{\uu}{{\bf u}}
\newcommand{\uua}{{\bf u}^{\alpha}}

\newcommand{\uai}{u^{\alpha} _i}
\newcommand{\uaj}{u^{\alpha }_j}

\newcommand{\ma} {m^{\alpha}}

\newcommand{\mb} {m^{\beta}}

\newcommand{\sab}{\sigma_{\alpha\beta}}
\newcommand{\gab}{g_{\alpha\beta}}
\newcommand{\muab}{\mu_{\alpha\beta}}

\newcommand{\bk}{\hat{\bf s}}

\begin{document}
\date{\today}
\title{Tracer diffusion of hard-sphere binary mixtures under nano-confinement}

\author{Umberto Marini Bettolo Marconi}
\email{umberto.marinibettolo@unicam.it}
%\address
\affiliation{ Scuola di Scienze e Tecnologie, 
Universit\`a di Camerino, Via Madonna delle Carceri, 62032 ,
Camerino, INFN Perugia, Italy}

\author{Paolo Malgaretti}
\email{malgaretti@is.mpg.de}
\affiliation{Max-Planck-Institut f\"{u}r Intelligente Systeme, Heisenbergstr. 3 D-70569 Stuttgart Germany}
\affiliation{IV. Institut f\"{u}r  Theoretische Physik, Universit\"{a}t Stuttgart, Pfaffenwaldring 57, D-70569 Stuttgart, Germany}
\affiliation{Departament de Fisica Fonamental, Universitat de Barcelona Barcelona, Av. Mart\'{i} i Franques 1, Barcelona, Spain}

\author{Ignacio Pagonabarraga}
\email{ipagonabarraga@ub.edu}
\affiliation{Departament de Fisica Fonamental, Universitat de Barcelona Barcelona, Av. Mart\'{i} i Franques 1, Barcelona, Spain}

\begin{abstract}
The physics of diffusion phenomena in nano and micro channels has attracted a lot of
attention in recent years, due to its close connection with many technological, medical and industrial
applications.
In the present paper we employ a kinetic approach to investigate how the confinement in nanostructured geometries 
affects the diffusive properties of fluid mixtures and leads to the appearance of properties different from those of bulk systems.
In particular, we derive an expression for
the  friction tensor 
in the case of a bulk fluid mixture
 confined to a narrow slit having undulated walls. The boundary roughness leads to a new mechanism for transverse diffusion, and can even lead to an effective diffusion along the channel larger than the one corresponding to a planar channel of equivalent section.
Finally we discuss a reduction of the previous equation to a one dimensional effective diffusion equation in which an entropic term encapsulates the geometrical information on the channel shape.
\end{abstract}

\maketitle

\section{Introduction}

The recent  interest for the transport phenomena of gases or liquids  confined in spaces whose span is comparable to the molecular size is motivated by the important technological, medical and industrial applications of nanofluidics so diverse as DNA sequencing, element separation, or energy harvesting. 
It is well known that confinement can have a strong impact on both the static and dynamic properties of fluids~\cite{bocquet2010nanofluidics,schoch2008transport}. The systematic theoretical study that confinement has on the behavior of fluids has concentrated on  straight channels, hence overlooking the potential impact that  the shape of the bounding walls may have on the collective properties of  confined fluids.
Recent studies have shown that the coupling between the system and the geometrical constraints overimposed by the environment can be relevant in situations such as molecular transport in zeolites~\cite{zeolites}, ionic channels~\cite{Calero2011}, or in microfluidic devices, ~\cite{Bezrukov,Fujita}. Moreover, geometrical constraints can induce novel dynamical scenarios, such as particle separation~\cite{Reguera2012}, cooperative rectification~\cite{Malgaretti2012,malgaretti2013confined}, negative mobility~\cite{Eichhorn2010,PaoloElecotrokinetics} that are absent in the behavior of the corresponding systems in bulk.

In the present paper we will analyze the impact that the corrugation of the confining walls has in the diffusion of model non-ideal fluids. Specifically, we will concentrate on  a hard sphere binary fluid mixture composed by two components, say A and B,  with different sizes, and will consider the tracer limit of  larger component, B.
We shall show that to take into account the wall roughness one has to modify the Fick equation for the concentration of B particles, $c(\rr,t)$:
$$
\frac{\partial}{\partial t} c(\rr,t) = D_0\nabla^2 c(\rr,t)
\label{eq:1}
$$
where $D_0$ is the  bulk diffusion coefficient.

Our treatment  considers a binary hard sphere mixture between hard walls, perhaps the most basic model of confinement of non ideal gases, which has the ability to capture the essential physics of inhomogeneous fluids, such as  variations of the density near the walls, layering and solid-fluid transition. 
We associate the majority component  with the solvent and the minority component with the solute, whose concentration is negligible,  the so called tracer limit and base the description of the diffusion process on the  equations for the partial densities and the momentum density, which have been previously obtained using a Boltzmann-Enskog approach \cite{marconi2011dynamics},
which  accounts for the excluded volume effect, due to the finite size of the molecules.
% and leads to a non trivial dependence of the matrix elements of the friction tensor  on the local densities of the species. 
In the tracer limit, since the majority species is virtually unaffected by the motion of the dilute species, we are able to derive  a simplified equation for the tracer concentration.
  
In the second part of the paper  following a seminal idea of Jacobs~\cite{jacobs} we further reduce the the complexity of the problem, by contracting the description from a  three dimensional problem, to a one dimensional effective problem, which is mapped onto a  diffusion process along the pore axis in the presence of a so called entropic potential. 
Such a reduction was  introduced heuristically by Jacobs many years ago and revisited by Zwanzig, Kalinay-Percus, Reguera-Rubi, Dagdug-Bezrukov-Berezhkovskii and coworkers among others who gave a statistical mechanical foundation to it~\cite{zwanzig,Reguera2001,Kalinay2006,kalinay2008,kalinay2010,kalinay2013,berezhkovskii2007diffusion,Berezhkovskii2011,dagdug2013,dagdug2013-2}. %{\bf IGN: An alternative to mention all these names is to say: "by Jacobs and has been revisited by a number of authors to provide a statistical mechanical foundation.}

Finally the magnitude of the coupling between the microscopic interactions of the medium and the mesoscopic modulation of the confinement are discussed.

%If the fluid density is low the interactions with the walls are dominant with respect to the those among the fluid-particles
%and there exists a number of studies on the subject, the so called Knudsen regime [CITE].

\section{The kinetic approach}
%%%%%%%%%%%%%%%%%%%%%%
In a recent series of papers one of the authors has developed a Boltzmann-Enskog self-consistent theory 
for fluid mixtures in inhomogeneous environment \cite{marconi2011dynamics,marconi2011multicomponent}
and has provided a microscopic derivation of  the equations  governing 
the evolution of concentration fluctuations in an M-component mixture of hard-spheres of diameters $\sigma_{\alpha\alpha}$
and masses $\ma$,  where $\alpha=1,M$.
The analysis of the kinetic equations  for the individual species leads to  the conservation law of the local number density of species, $\na(\rr,t)$:  
\be
\frac{\partial}{\partial t} \na(\rr,t) +\nabla\cdot \Bigl( \na(\rr,t) \uua(\rr,t)\Bigl)=0 .
\label{continuity}
\ee
and to the balance equation for
the associated momentum density:
\begin{widetext}
\bea
&&
\frac{\partial}{\partial t} [\na(\rr,t)\uaj(\rr,t)]+ 
\nabla_i \Bigl(\na(\rr,t) \uai(\rr,t) \uaj(\rr,t)
- \na(\rr,t) w^{\alpha}_i(\rr,t) w^{\alpha}_j(\rr,t)\Bigl)=
\nonumber\\
&& 
-\frac{1}{\ma}\nabla_i \pi_{ij}^{\alpha}- \frac{\nabla_j V^{\alpha}(\rr)}
{\ma}\na(\rr,t)+\frac{\na(\rr,t)}{\ma}\Bigl( \FF^{\alpha,mf}(\rr,t)+\FF^{\alpha,drag}(\rr,t)+\FF^{\alpha,viscous}(\rr,t) \Bigl) .
\label{momentpartial}
\eea
\end{widetext}

%[SEGNO SBAGLIATO di V]
 where  $\uua$ is the average velocity of component $\alpha$, $\bf{w}^\alpha=(\uua-\uu)$
 is the relative average velocity of species $\alpha$ with respect to the center of mass  velocity, $\uu$, of the mixture. 
 
 The tensor $\pi_{ij}^{\alpha}(\rr,t)$, in  analogy with pure fluids, represents the kinetic contribution to the partial stress tensor
 \be
\pi_{ij}^{\alpha}(\rr,t)=  k_B T  \delta_{ij} \na(\rr,t)
\label{pressurekin}
\ee
Eq.~\eqref{momentpartial}, besides the external body force $-\nabla_j V^{\alpha}$, 
contains three kinds of forces of different nature,
 resulting from the analysis of the microscopic Enskog collision operator \cite{van1973modified,van1973modified2}:
 a mean force, a drag force  and a viscous force. Specifically,
 $\FF^{\alpha,mf}$ is the reversible force  acting on $\alpha$ particles at $\rr$ due to
the influence of all remaining particles: 
\begin{multline}
\FF^{\alpha,mf}(\rr,t)=-k_B T\\
\sum_\beta\sab^2 
\int d\bk \bk
g_{\alpha\beta}(\rr,\rr+\sab\bk,t)
n_{\beta}(\rr+\sab\bk,t)
\end{multline}
% \be
% \FF^{\alpha,mf}(\rr,t)=-k_B T\sum_\beta\sab^2 
% \int d\bk \bk
% g_{\alpha\beta}(\rr,\rr+\sab\bk,t)
% n_{\beta}(\rr+\sab\bk,t)
% \ee
where
 $\bk$ is a unit vector of components $(\sin\theta\cos\phi,\sin\theta\sin\phi,\cos\theta)$, $\int d\bk \equiv
\int_0^{2\pi} d\phi \int_0^\pi d\theta\sin\theta$ 
 indicates integration over the unit sphere, 
 $g_{\alpha\beta}$ is the inhomogeneous hard sphere pair correlation function
 at contact and $\sab=(\sigma_{\alpha\alpha}+\sigma_{\beta\beta})/2$.
Such a force, is the gradient of  the so-called potential of mean force and can be identified
with the gradient of the excess chemical potential of species $\alpha$ over the ideal gas  value, $\mu_{int}^{\alpha}(\rr)$
according to:
\be
\FF^{\alpha,mf}(\rr,t)= -\NN \mu_{int}^{\alpha}(\rr,t).
\label{potchimico}
\ee
whereas the corresponding  total chemical potential is given by $\mu^\alpha(\rr)=k_B T \ln \na(\rr)+\mu_{int}^{\alpha}(\rr)$.
The  drag force is purely dissipative, local and is proportional to  the difference of velocities of
unlike species:
\be
F_i^{\alpha,drag}(\rr,t)= 
-\sum_\beta{\bf \gamma}_{ij}^{\alpha\beta} (\rr,t)  (u^\alpha_j(\rr,t)-u^\beta_j(\rr,t))
\label{gammaab}
\ee
via the inhomogeneous friction tensor $ \gamma_{ij}^{\alpha\beta}$ which is associated to the interactions and relative motion between the different species.
Finally, the dissipative force  $\FF^{\alpha,viscous}$ represents the viscous force acting on species $\alpha$ 
due to  velocity gradients and in the present treatment it will be neglected 
under the assumption  that the velocity varies slowly  \cite{marconi2009kinetic}.

The  details of the procedure which allows to reduce
the coupled differential equations for the densities and for the momenta of the species to a diffusion
equation for the concentration  are described in refs. \cite{marconi2011dynamics,marconi2011multicomponent}. 
The assumptions are: a) 
that, since the typical fluid velocities 
 in micro and nanofluidic systems  are  low,  we can neglect the non linear terms, 
 b) the velocity gradients are negligible so that we can discard viscous forces and
 retain only the diffusive terms, and 
 c)   the acceleration of the species with respect to the center of mass is negligible.

For hard sphere mixtures the friction tensor in the Enskog model  can be computed explicitly and reads:
\begin{multline}
\gamma_{ij}^{\alpha\beta}(\rr,t)=2\sigma_{\alpha\beta}^2 \sqrt{\frac{2\muab k_B T}{\pi} }\\
\int d\bk s_i s_j
\gab(\rr,\rr+\sigma_{\alpha\beta}\bk,t)
\nb(\rr+\sigma_{\alpha\beta}\bk,t),
\label{tensorgamma}
\end{multline}

where $\muab =\ma \mb/(\ma+\mb)$ is
the reduced mass.
 
 %%%%%%%%

%%
%%
% OFF DIAGONAL
%\subsection{Off diagonal friction tensor elements. X parallel to walls. Z normal to walls}

%\subsection{Bulk concentration diffusion}

For  a binary mixture of hard-spheres it is useful to  define
the local mass concentration
$c(\rr,t)=\frac{\rho^B(\rr,t)} {\rho(\rr,t) }$, with $\rho^B=m^B n^B$ and $\rho=m^A n^a+m^B n^B$.
Using the results of Ref.~\cite{marconi2011multicomponent}, when the fluid velocity is negligible and in the absence of external forces, the balance equation  for the local mass concentration reduces to
\begin{multline}
\frac{\partial}{\partial t} c(\rr,t) =\frac{1}{\rho(\rr,t)}
\nabla_i\cdot \Bigl(\rho(\rr,t) c(\rr,t)(1-c(\rr,t))
 (\gamma_{ij}^{-1}(\rr,t))\\
  \nabla_j ( \frac{\mu^{B}(\rr,t)} {m^B}- \frac{\mu^{A}(\rr,t)}{m^A})\Bigl) ,
\label{continuityconcentration2}
\end{multline}

where we have introduced the symmetrized inhomogeneous friction tensor, $\gamma_{ij}=\frac{\gamma_{ij}^{AB}}{m^A}+\frac{\gamma_{ij}^{BA}}{m^B}$ and its inverse $\gamma_{ij}^{-1}$. Neglecting the variations in the density around a reference  sphere the friction tensor reduces to $\gamma_{ij}=\gamma \delta_{ij}$. We can then derive an approximate, explicit expression for $\gamma$ from Eq.~\eqref{tensorgamma},
%In the case of uniform densities using eq.\eqref{tensorgamma}  we obtain a  constant friction coefficient
\be
\gamma=\frac{8}{3}\rho\frac{\sqrt{2\pi\mu_{AB} k_B T }}{m^A m^B} g_{AB} \sigma_{AB}^2,
\label{gammasymmetrized1}
\ee
and rewrite the conservation equation for the local  mass concentration as a  standard diffusion equation
\be
\frac{\partial}{\partial t} c(\rr,t) = D^{AB} \nabla^2 c(\rr,t),
\label{continuityconcentration2-1}
\ee
in terms of  the mutual diffusion coefficient, $D^{AB}$ ,
\be
D^{AB} 
= \frac{  k_B T}{\gamma}\frac{\rho}{n}\frac{1}{m^A m^B},
\label{dab}
\ee
which relates the friction  and the diffusion coefficients through
the Einstein fluctuation-dissipation  relation.

\section{Diffusion in confined structures}
%\subsection{Si passa ad altro}

The general framework described in the previous section also holds for  a nanoconfined  mixture . To analyze its dynamic features, we consider the specific case where the fluid is confined  to a  symmetric slit bound by two non intersecting  walls identified, in the Monge representation, by the  two height functions $z = \pm  h(x)$. Accordingly, the fluid densities will be symmetric   with respect to the midplane $z=0$, and translationally invariant along the y direction.
Each species, $\alpha$, composing the  fluid  mixture is confined to the  slit due to the wall potential they will experience,

%  To investigate how the
% confinement of the fluid in nanoscopic systems affects  the  diffusive behavior we consider the friction matrix \eqref{tensorgamma}. This quantity
% deviates from  its bulk value in proximity of  the 
% boundaries, since
% molecules adjacent the walls experience a highly inhomogeneous environment.
% 
% We model the channel by means of a slit bounded by  two non intersecting surfaces identified in the Monge representation by two height functions $z=\pm h(x)$, where  $z$ is the distance measured with respect to the symmetry plane at $z=0$. For the sake of simplicity we assume reflection symmetry about the midplane $z=0$   and translational symmetry along the $y$ direction so that the hydrodynamic variables depend only on the $x$ and $z$ coordinates.
%  
 %The particles of type $\alpha$ experience the following wall-potential:
 \begin{equation}
  V^\alpha(\rr)=
  \begin{cases}
   V^\alpha_{soft}(\rr) & z<|h(x)|\\
   \infty & else
   \end{cases}
   \label{eq:V_soft}
 \end{equation}
that is the sum of a soft attractive potential, $V^\alpha_{soft}$, of general functional form  and of a harshly repulsive confining potential.
The densities of the fluid species are no longer homogeneous due to the presence of the confining solid walls. As a result, the friction matrix, Eq.~\eqref{tensorgamma}, is no longer diagonal. The inhomogeneous nature of the pair correlation function  at contact, $g_{AB}$, encodes the response of the mixture to such inhomogeneities. Although an exact functional form of the pair distribution function under generic  inhomogeneous conditions is not known, we resort to a two-component generalization~\cite{wendland1997born} of the Fisher-Methfessel prescription~\cite{fischer1980born}, which states that the 
%  
% The calculation of the friction tensor \eqref{tensorgamma} involves
% the  inhomogeneous pair correlation function, $g_{AB}$,  at contact whose form is  
% not exactly known, so that we have to resort to
% a two-component generalization \cite{wendland1997born} of the Fischer-Methfessel prescription\cite{fischer1980born}
% stating that the 
functional form of the  inhomogeneous  $g_{AB}(\rr,\rr+\sigma_{AB}\bk)$ can be obtained from
the  Carnahan-Starling expression for the bulk pair correlation  of mixtures at pair contact, $g_{AB}^{bulk}$,   \cite{boublik1970hard,mansoori1971equilibrium} 
\begin{multline}
g_{AB}^{bulk}(\{ \xi_n\})
=\frac{1}{1-\xi_3}+\frac{3}{2}
\frac{\sigma_{AA}\sigma_{BB}}{\sigma_{AB}}\frac{\xi_2}{(1-\xi_3)^2}
+\\
+\frac{1}{2} \Bigl(\frac{\sigma_{AA}\sigma_{BB}}{\sigma_{AB}}\Bigl)^2
\frac{\xi_2^2}{(1-\xi_3)^3}.
\label{carnahan}
\end{multline}

where
 the  $ \xi_n$ are linear combinations of the bulk densities
$
 \xi_n =\frac{\pi}{6}\sum_{\alpha} n^\alpha \sigma_{\alpha\alpha}^n \, .
$
In an inhomogeneous environment $g_{AB}$ is generalized replacing the bulk densities $n^A$ and $n^B$  by the corresponding inhomogeneous coarse grained  densities
$\bar n^A(\rr)$ and $\bar n^B(\rr)$. These are the averages 
over spheres of volume
$\omega_{\alpha}=\pi \sigma_{\alpha\alpha}^3/6$ centered at $\rr$:
%associated to the coarse grained density,
$$
\bar n^\alpha(\rr)=\frac{1}{\omega_\alpha}\int d\rr'
n^\alpha(\rr+\rr')\, \theta\left( \frac{\sigma_{\alpha\alpha}}{2}-|\rr-\rr'|\right)
$$
Accordingly, we assume that the spatial dependence of the pair correlation function at contact enters through its dependence on the inhomogeneous coarse grained densities. Hence, from Eq.~\eqref{carnahan} we arrive at
\be
g_{AB}(\rr,\rr+\sigma_{AB}\bk)
=g_{AB}^{bulk}(\{\bar\xi_n(\rr+\frac{1}{2}\hat{\bf s} \sigma_{AB})\})
\label{pair_corr_contact}
\ee
where the smeared functions $\bar \xi_n$ are
$$
\bar\xi_n(\rr) =\frac{\pi}{6} \bar n^A
(\rr) \sigma_{AA}^3+
\frac{\pi}{6} \bar n^B(\rr) \sigma_{BB}^3 .
$$
Substituting Eq.\eqref{pair_corr_contact} into Eq.\eqref{tensorgamma} we get:
\begin{multline}
\left( \begin{array}{cc}
  \gamma_{XX}(x,z) \\
\gamma_{ZZ}(x,z)\\
\gamma_{XZ}(x,z)
 \end{array} \right)
 =\frac{2\sigma_{AB}^2}{m^B} \sqrt{\frac{2\mu_{AB} k_B T}{\pi} }\\
\int d\bk \Biggl\{\left( \begin{array}{cc}
  s_X s_X \\
s_Z s_Z\\
s_X s_Z
 \end{array} \right)
 g_{AB}^{bulk}(\{\bar\xi_n(\rr+\frac{\hat{\bf s} \sigma_{AB}}{2})\})\\
 [\frac{n^A(\rr+\hat{\bf s} \sigma_{AB})} {m^A}+\frac{n^B(\rr+\hat{\bf s} \sigma_{AB}) }{m^B}]\Biggl\}.
\label{tensorgammabb-paolo}
\end{multline}
which shows that the geometric confinement leads, generically, to a tensorial friction coefficient that deviates from its bulk behavior. As we will analyze subsequently, the geometrically-induced off-diagonal components imply that  in general, under confinement, non-ideal fluid mixtures will show transverse diffusion 
These off-diagonal components vanish for planar interfaces, hence showing that transverse diffusion develops only as a combination of particle interaction and spatially varying  confinement. Specifically, if we consider flat walls, Eq.\eqref{tensorgammabb-paolo} reduces to 
\begin{multline}
\left( \begin{array}{cc}
  \gamma_{XX}(x,z) \\
\gamma_{ZZ}(x,z)\\
 \end{array} \right)
=
\frac{2\sigma_{AB}^2}{m^B}   \sqrt{ 2 \pi\mu_{AB} k_B T } \\
\int_{-1}^1 d \hat s_Z  \Biggl\{
\left( \begin{array}{cc}
 1-\hat s_Z^2 \\
 2 \hat s_Z^2 \
 \end{array} \right)
 g_{AB}^{bulk}(\{\bar\xi_n(z+\frac{1}{2}\sigma_{AB} \hat s_Z)\})\\
 [\frac{
n^A(z+\sigma_{AB} \hat s_Z)}{m^B}+
\frac{
n^B(z+\sigma_{AB} \hat s_Z)}{m^A}]\Biggl\}
\label{normalc}
\end{multline}
These expressions are equivalent to those obtained by Davis and coworkers for slit-like pores by 
using a method based on the Born-Green-Yvon equation  \cite{vanderlick1987self,davis1987kinetic,bitsanis1987molecular,magda1985molecular}.

\section{Diffusion equation for $n^B$ in the tracer limit}

% CONTINUITY CONCENTRATION
The results obtained in the previous section holds for a general binary mixture of hard spheres.
When one of the components of the binary mixture becomes extremely diluted
($n^B<< n^A$) the method described in the previous sections to derive the corresponding diffusion equations for the mixture dynamics considerably simplifies and one can arrive at more explicit expressions for the diffusive fluxes
In the tracer limit, in fact, we may disregard the B-B interactions and the friction matrix governing the diffusion of B particles turns out to be independent from the density profile of the B-particles and is solely a functional  of the solvent density profile, $n^A(\rr,t)$.
In the following we treat the case of a system prepared in a state where the  majority component A, identified with the solvent, is at thermodynamic equilibrium, so that $\uu=0$ and its density satisfies the balance condition:
\be
\nabla (\mu^A(\rr)+V^A(\rr)) =0 .
\label{bequil}
\ee
Although it will not be pursued in the rest of this contribution, the  method we will introduce can be easily generalized to non-equilibrium situations where the dynamics of A particle is known and independent of the B particles.

If we specialize the general Eq.~\eqref{eq:1} for the tracer species, with density $n^B$, and use Eq.~\eqref{continuity} to relate the velocity of the tracer species to local forces we arrive at 
\begin{multline}
\frac{\partial}{\partial t} n^B(\rr,t)=\nabla_i\cdot \Bigl(n^B(\rr,t)\gamma^{-1}_{ij}(\rr,t)\Big[   \frac{1}{m^B}\nabla_j  ( \mu^B(\rr)+\\
+V^{B}(\rr)) -  \frac{1}{m^A}\nabla_j  (\mu^A(\rr)+V^{A}(\rr))\Big]   \Bigl)
\label{continuityrhob2}
\end{multline}
where in this limit $\gamma^{-1}_{ij}$ reduces to the  $ij$ matrix element of the matrix inverse of  $\gamma_{ij}= \frac{1}{m^B}\gamma_{ij}^{BA}$ .

For  the confining slit, Eq.~\eqref{eq:V_soft}, which is translational invariant along the y-direction,  the relevant components of the  flux of the tracer species in response to density gradients or externally applied forces, read
\begin{widetext}

%In the case of a two dimensional motion occurring  in response to gradients in density or forces only in the directions $x,z$ we find that the relevant components of the diffusive flux are given by
\be
J^B_X(x,z)=-\frac{n^B(x,z,t)}{m_B} \Bigl( \gamma^{-1}_{XX}(x,z) (\nabla_X \mu^B(x,z,t)+\nabla_X V^{B}(\rr))+
\gamma^{-1}_{XZ}(x,z) (\nabla_Z \mu^B(x,z,t)+\nabla_Z V^{B}(\rr))  \Bigl)
\label{ccp-x}
\ee
and
\be
J^B_Z(x,z)=-\frac{n^B(x,z,t)}{m_B} \Bigl(  \gamma^{-1}_{ZX}(x,z) (\nabla_X \mu^B(x,z,t)+\nabla_X V^{B}(\rr)) 
+\gamma^{-1}_{ZZ}(x,z) (\nabla_Z \mu^B(x,z,t)+\nabla_Z V^{B}(\rr)) \Bigl)
\label{ccp-z}\ee
\end{widetext}
The inhomogeneous confinement  keeps the tensorial structure of the friction of the tracer species due to its interaction with the majority species, A. The structure of the fluxes depend only on  chemical potential gradients in the tracer species and in the direct effect of external fields on them. However, the friction coefficients depend functionally on the  equilibrium, inhomogeneous, profiles of the majority species, $n^A(r)$, in the slit. In particular, the knowledge of $n^A(\rr)$ also allows, as
shown in the following, to compute the non uniform 
contact value of the pair correlation $g^{AB}$ which is needed to evaluate
the friction matrix.
The structure of $\gamma_{XZ}$ shows that only near a non planar substrate this matrix element is non vanishing. 
In order to give an estimate of this quantity we perform a Taylor expansion of the density and of the pair correlation
in powers of  the displacement $\sigma_{AB}\bk$ up to second order
\begin{multline}
\gamma_{XZ}(x,z)\approx \frac{8 \sigma_{AB}^4}{ 15 m^B} \sqrt{2\pi \mu_{AB} k_B T}\cdot \\
\cdot\frac{\partial^2 }{\partial x \partial z}[n^A(x,z) g_{AB}^{bulk}(\bar\eta(x,z))] 
\label{eq:gammasimplifyed}
\end{multline}
and observe that for the appearance of  non diagonal friction tensor elements it is necessary to have non vanishing
cross  derivatives $\frac{\partial^{m+n} n^A(x,z)g_{AB}^{bulk}(\bar\eta(x,z))}{\partial^m x \partial^n z}\neq 0$, with $m,n$ odd integers . These terms vanish as we move away from the interfaces, where the $n^A$ profile becomes nearly constant. Moreover, to zeroth order in the gradients of the density of the A particles the off-diagonal matrix element vanishes because of the parity of the integrals. 

The present treatment generalizes the method proposed by Davis and coworkers to non flat confining surfaces and  allows to treat also mixtures with finite concentrations of host particles. However, in this case the friction matrix is time dependent since the motion of the $B$ particles affects the configuration of the majority species ($A$ particles). Moreover, also the interactions among the B particles would give a finite contribution to the friction matrix \cite{marconi2011multicomponent} . 

The off-diagonal component of the friction matrix due to the inhomogeneous confinement affects the overall longitudinal diffusion of the tracers along the channel. From Eqs.~\eqref{ccp-x} and~\eqref{ccp-z}, in the absence of a net transverse flux, $J_Z=0$,  the response to an applied force along the channel (or equivalently the relaxation of an equilibrium fluctuation) is  characterized by the effective longitudinal diffusion coefficient, 

% The presence of off-diagonal term in Eq.\eqref{tensorgammabb} affects the overall diffusion along the x-direction
% as one can see  
% introducing an appropriate longitudinal diffusion coefficient according to the formula

\be
D_{XX}(x,z)=\frac{k_B T}{m^B}\gamma^{-1}_{XX}(x,z)
\label{eq:diff1}
\ee
which has the explicit form:
\begin{multline}
D_{XX}(x,z)=\frac{\gamma_{ZZ}(x,z)} {\gamma_{XX}(x,z) \gamma_{ZZ}(x,z)-\gamma_{XZ}(x,z)^2}=\\
=\frac{1}{\gamma_{XX}(x,z)}\left[ 1-\frac{\gamma_{XZ}(x,z)^2}{\gamma_{XX}(x,z) \gamma_{ZZ}(x,z)}\right]^{-1} .
\label{eq:DXX-expansion1}
\end{multline}
Eq.\eqref{eq:DXX-expansion1} shows that, due the positivity of the matrix $\gamma_{ij}$, the net effect of the off-diagonal terms is to enhance the diffusivity along  the longitudinal direction. The magnitude of such an enhancement depends on that of the off-diagonal term in Eq.\eqref{tensorgammabb-paolo}. In order to characterize the magnitude of such contributions we use Eq.\eqref{eq:gammasimplifyed} into \eqref{eq:DXX-expansion1}   getting:
\be
\frac{\gamma_{XZ}^2(x,y)}{\gamma_{XX}(x,y)\gamma_{ZZ}(x,y)}\simeq\left(\frac{\sigma^2_{AB}}{Lh}\frac{n^Am^A}{\rho}\right)^2.
\ee
The last expression shows that if the particle-particle distance at contact, $\sigma_{AB}$, is smaller than the channel 
width $2h$ and  the channel  varies smoothly, $h\ll L$ (so that $\frac{\gamma_{XZ}^2(x,y)}{\gamma_{XX}(x,y)\gamma_{ZZ}(x,y)}\ll 1$)  we can expand Eq.\eqref{eq:DXX-expansion1} obtaining: 

\begin{multline}
D_{XX}(x,z)\simeq\frac{1}{\gamma_{XX}(x,z)}\left[ 1+\frac{\gamma_{XZ}(x,z)^2}{\gamma_{XX}(x,z) \gamma_{ZZ}(x,z)}\right]\simeq\\
\simeq\frac{1}{\gamma_{XX}(x,z)}\left[1+\left(\frac{\sigma^2_{AB}}{Lh}\frac{n^Am^A}{\rho}\right)^2\right]
\label{eq:DXX-expansion2}
\end{multline}
that captures the translational diffusivity enhancement along the longitudinal direction. The second  relation in Eq.\eqref{eq:DXX-expansion2} provides an order of magnitude estimate of the impact that channel corrugation has on the enhancement of longitudinal diffusion. 
% Expression  \eqref{eq:DXX-expansion2} clearly indicates that under the hypothesis above the diffusion coefficient  is  enhanced by the presence of modulations in the channel amplitude.
Hence the time evolution of distribution of particles in an inhomogeneous environment will experience a faster broadening as compared to the case of the same distribution in an homogeneous environment.
 It is interesting to note that such a result can be extended to other scenarios where the off-diagonal terms in $\gamma_{ij}$  have a different physical origin, {\sl e.g.}Êwhen they are induced by  spatial variations of an imposed potential. 
 %%%%%%%%
 
 \section{Effective one dimensional diffusion equation a la Fick-Jacobs}

%\section{One dimensional reduction}
%In the section above we have derived and computed the diffusion matrix elements in the case of a non uniform slit.
The diffusion equation for the tracer species in the corrugated channel, Eq.~\eqref{continuityrhob2}, requires solving a two dimensional elliptic partial differential equation which can be  numerically involved. In fact, even  the determination of 
the steady current ,${\bf J}^B(x,z)$, generated by an externally imposed gradient of concentration or by a force field can become numerically intensive.
When the corrugation of the walls is slowly varying  one can  further approximate  the diffusion process and obtain a mathematically simpler  description following the arguments originally proposed by Jacobs~\cite{jacobs}.
Here we shall use a method closely related to the one introduced by Zwanzig~\cite{zwanzig1992diffusion} which eliminates  the transverse degrees of freedom  assuming a fast equilibration of the density profile in the transverse
direction, $z$~\cite{marconi2013effective}. The validity of the resulting  effective one dimensional equation, referred to as Fick-Jacobs equation, has recently been confirmed~\cite{Reguera2006,Burada2007,kalinay2006corrections,kalinay2008,martens2011entropic} %by the asymptotic analysis of Martens {\sl et al.}
, who also considered  the functional forms of higher order correction to the effective one dimensional dynamics.

If the channel width varies slowly, $\nabla_x h<<1$, one can assume that the tracer distribution equilibrates
on a time scale faster than the one needed to reach equilibrium along the direction of the variation of the
channel section. 
%When the tracer distribution reaches its equilibrium value along the transverse direction one has
%{\bf PM: In the case of vanishing flux along the normal direction at the channel walls (see Appendix A)}
Accordingly, the transverse equilibrium  condition for the B-particles reads
\be
 \nabla_Z( \mu^B(x,z)+ V^{B}(x,z))=0
 \label{trasvequil}
 \ee
 and the continuity Eq.~\eqref{continuityrhob2}
 can be reduced to 
 \begin{multline}
\frac{\partial n^B(x,z,t)}{\partial t}=\frac{1}{k_B T}\Bigl[\nabla_ X \gamma^{-1}_{XX}(x,z) +\nabla_ Z\gamma^{-1}_{ZX}(x,z)
\Bigl]\cdot\\
\cdot\Bigl( n^B(x,z,t) \nabla_X \mu^B(x,z,t)+n ^B(x,z,t)\nabla_X V^{B}(\rr))\Bigl).
\label{final-eq}
\end{multline}
To use the information contained in Eq.~\eqref{trasvequil} we separate the local chemical potential $\mu^B$ into its ideal gas contribution and the interacting part:
\be
\mu^B(x,z,t)=k_B T \ln n^B(x,z,t)+\mu_{int}^B(x,z,t) \, ,
\ee
assuming that the tracer density can be decomposed in its concentration along the channel, $c(x,t)$, and the conditional density across the channel section, $P$
%make the following factorization ansatz for the one particle distribution function of tracers:
%and substituting in \eqref{trasvequil},  we obtain:
\be
n^B(x,z,t)=c(x,t) P(z|x).
\label{Zwanzig}
\ee
and substitute this expression  in Eq. \eqref{trasvequil}.
The conditional  density profile, $P(z|x)$, is obtained from the equilibrium condition in the transverse direction, for any choice, of $c(x,t)$:
\be
P(z|x)=\,\frac{e^{-\beta(\mu^B_{int}(x,z)+ V^B(x,z))}} { \int_{-\infty}^{\infty} dz e^{-\beta(\mu^B_{int}(x,z)+ V^B(x,z))}} \, .
%P(z|x) &=&0 \,\,\, |z|> h(x)
\label{Zwanzig3b}
\ee
This factorization led Zwanzig to generalize the Fick-Jacobs equation for a two-dimensional potential for non-interacting diffusing particles~\cite{zwanzig1992diffusion}. Deviations from this conditional local equilibrium can be included still in an effective, generalized Fick-Jacobs equation where the diffusion coefficient depends on the channel shape~\cite{zwanzig1992diffusion,Berezhkovskii2011,martens2011entropic}. For interacting systems, and following Zwanzig,
%was shown by According to Zwanzig  \cite{zwanzig1992diffusion},
one interprets $P(z|x)$ 
as the local equilibrium distribution of $z$  conditional on a given $x$. For simplicity, we will always assume that the local equilibrium factorization, Eq.(~\ref{Zwanzig}), holds. Accordingly, the results we will derive will generalize the original Fick-Jacobs equation for interacting systems. It is worth stressing that
%Let us remark  that 
in the tracer limit,  $\mu^B_{int}(x,z) $ only depends on $n^A$, but not on $n^B$ so that it is a quenched
variable.

The factorization assumption, Eq.~\eqref{Zwanzig}, allows us to arrive at a simplified expression for the diffusive fluxes of tracer particles, whose component along the channel reads
%We now substitute $n^B$ using \eqref{Zwanzig} into \eqref{ccp-x}  and obtain the tangential component of the current:
\begin{multline}
J^B_X(x,z)=-\frac{c(x,t)}{k_B T}  P(z|x)  \gamma^{-1}_{XX}(x,z)\Bigl( \frac {k_B T }{c(x,t)} \nabla_X c(x,t)+\\
+\frac {k_B T }{P(z|x)} \nabla_X P(z|x)+\nabla_X \mu^B_{int}(x,z) +\nabla_X V^{B}(x,z) \Bigl)
\label{ccp-xd}
\end{multline}
while its   normal component can be expressed as
\begin{multline}
J^B_Z(x,z)=-\frac{c(x,t)}{k_B T}  P(z|x)  \gamma^{-1}_{ZX}(x,z)\Bigl( \frac {k_B T }{c(x,t)} \nabla_X c(x,t)+\\
+\frac {k_B T }{P(z|x)} \nabla_X P(z|x)+\nabla_X \mu^B_{int}(x,z) +\nabla_X V^{B}(x,z) \Bigl)
\label{ccp-xe}
\end{multline}
The homogenization approximation implies that both flux  components depend only  on the concentration gradients along the channel. The inhomogeneous nature of the channel is encoded in the effective friction matrices  and the expressions for the density profits and excess chemical potentials.

The net particle flow along the channel , $I_x$, can be obtained averaging the longitudinal component over the local channel section for no-flux boundary  conditions across the channel walls (see Appendix A)
\begin{multline}
I_X(x)=\int dz J^B_X(x,z)=\\
=-\mathcal{D}_{XX}(x) \,\Bigl( \nabla_X c(x,t)+c(x,t)\beta \nabla_X \mathcal{A}(x)  \Bigl)
\label{jbm2}
\end{multline}
where
\be
 \beta \mathcal{A}(x)  =-\ln\Bigl(\frac{1}{L}
 \int_{-\infty}^{\infty} dz e^{-\beta(\mu^B_{int}(x,z)+ V^B(x,z))}\Bigr)
\label{mu0}
\ee
is the effective potential accounting for both the confinement and the interactions with the $A$ particles encoded in $\mu_{int}$.  Interestingly, $\mathcal{A}$ can be regarded as a free energy, being the logarithm of the integral of the Boltzmann factor. Although we have introduced the prefactor $\frac{1}{L}$ in Eq.~\eqref{mu0} for dimensional reasons, its magnitude is irrelevant in the subsequent analysis since only derivatives of $\mathcal{A}$ affect particle dynamics. In Eq.~\eqref{jbm2} we have introduced the pore averaged diffusion , $\mathcal{D}_{XX}$, one  component of the pore averaged diffusion matrix
\be
\mathcal{D}_{ij}(x)=
\int_{-\infty}^{\infty} dz P(z|x) |D_{ij}(x,z)|,
\label{dij}
\ee
which  is very similar to the  pore averaged  diffusivity  proposed by 
Davis and coworkers~\cite{davis1987kinetic}. % , who only considered the diagonal elements of the tensor.
Eq.~\eqref{jbm2} generalizes the Fick-Jacobs equation to systems of interacting particles in the tracer limit and, for non interacting particles, $\mu_{int}\rightarrow 0$, takes the form of the standard entropic potential contribution to particle transport along a corrugated channel, where the diffusion matrix is diagonal and with components that become constant. It is worth pointing out that  non-ideality leads to an effective  position-dependent local diffusion coefficient  from the homogenization process assuming that the density profiles can be factorized assuming local equilibrium, Eq.(~\ref{Zwanzig}). For ideal systems, a position-dependent diffusion coefficient arises from deviations in the density profile from  Eq.(~\ref{Zwanzig}). These changes have been obtained performing   higher-order perturbation theories from  the original Fick-Jacobs equation~\cite{zwanzig1992diffusion,Berezhkovskii2011,martens2011entropic}. This difference shows that the origin of the inhomogeneous effective diffusion after homogenization for interacting systems is qualitatively different from the effective, local diffusion derived  for ideal systems in the framework of the Fick-Jacobs equation. A generalization of the density profile beyond Eq.(~\ref{Zwanzig}) for non-ideal systems, outside the scope of this paper,  will clarify the relative relevance of the different contributions to a local, position-dependent, diffusion matrix for highly corrugated channels.

Eq.~(\ref{ccp-xd}) shows that the flux along the channel is only a function of the local $D_{XX}$ component of the  pore averaged diffusion matrix. However, the transverse flux, Eq.~(\ref{ccp-xe}) will depend on the cross component $D_{XZ}$, which does not vanish only for interacting systems in corrugated  channels.
%, in contrast to the case of non-interacting particles for which such  correction are introduced {\it a posteriori} in order to improve the approximation}   

Although the procedure described can be easily extended beyond the tracer limit, the results obtained are formal and more complex to use because the friction matrix depends in general on the concentration of both A and B particles, which depend now on time.
% On the other hand, going beyond the tracer limit is difficult  because the friction matrix would depend on the finite concentration of the $B$ particles and would acquire a time dependence. 
Notice that $V^B(x,z)$ and $\mu_{int}^B(x,z)$ are known because   in the tracer limit  depend only on the majority species
$A$, so that $P(x|z)$ is time independent. The current intensity  \eqref{jbm2} contains a one dimensional diffusive term, a drift term due to the interactions with the B particles and the walls and an entropic term due to the non uniform shape of the channel.

%%%%%%%%%%%%%%%%%%%

%\subsection{Fick-Jacobs equation for partilces moving in a porous medium}
Since at steady state $I_X$  must be constant on every section, due to particle number conservation, we can express the evolution equation for the tracer density,$c(x)$, simply as 

\be
 \frac {d c(x)}{dx}{\bf +}\beta \frac {d \mathcal{A}(x)}{dx} c(x)  =-\frac{I_X}{\mathcal{D}_{XX}(x)},
 \label{firstorderb}
  \ee
which can be expressed as a quadrature 
\be
c(x) = e^{-\beta \mathcal{A}(x)}\left [ -I_X \int_0^x \frac{e^{ \beta\mathcal{A}(s)}}{\mathcal{D}_{XX}(s)}ds + e^{\beta \mathcal{A}(0)} c(0) \right] .
 \label{integral}
\ee
%{\bf where $\Pi$ is a constant introduced in order to satisfy the boundary conditions.}
At equilibrium, since $I_X=0$, the tracer species steady state density profile reduces to its equilibrium shape
\be
c(x)=e^{-\beta \mathcal{A}(x)} e^{\beta \mathcal{A}(0)} c(0), 
\label{eq:c_x}
\ee
as expected.
%%%%%%%%%%%%%%
\begin{figure}[htb]
\includegraphics[scale=1.] {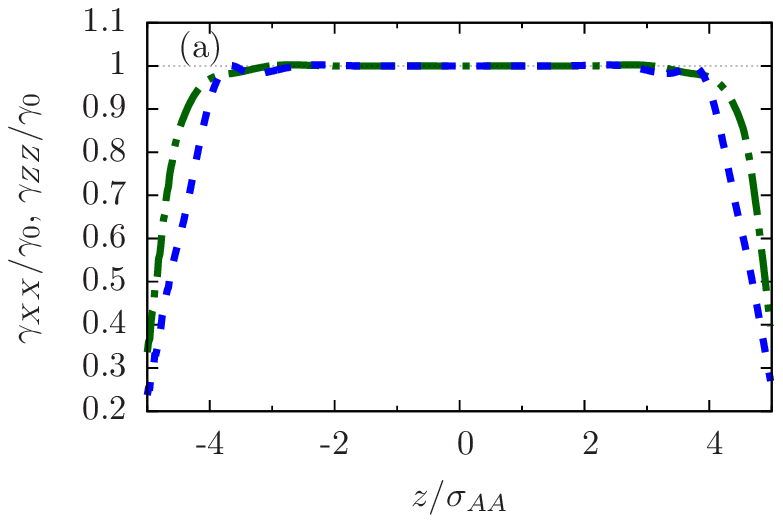}\\
\includegraphics[scale=1.] {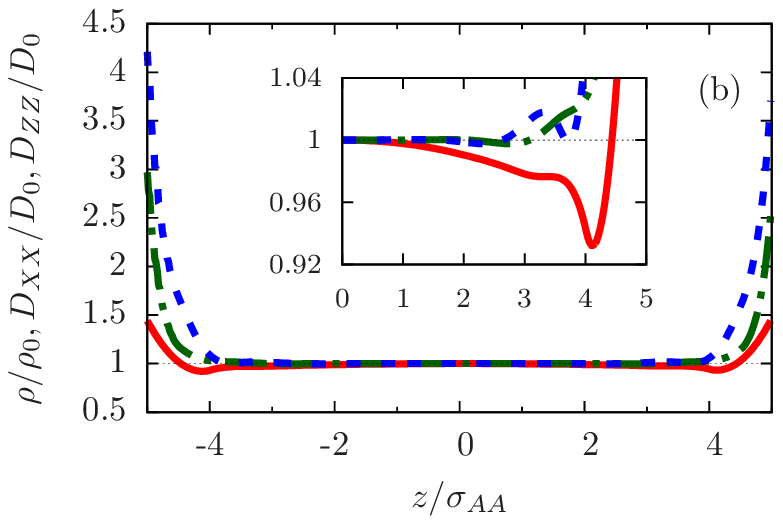}
\caption{(colors online) Flat walls. A: friction matrix elements $\gamma_{XX}$ (green dot-dashed line) and $\gamma_{ZZ}$ (blue dashed line) as a function of the distance from the center of the pore normalized by the particle size $\sigma_{AA}$. B: Density profiles (red solid line), normalized by its values at the center of the channel, $\rho_0=0.2$, and $D_{XX},D_{ZZ}$ (green dot-dashed and blue dashed line respectively) normalized by their values at the center of the channel as a function of the transverse coordinate, $z$, normalized by the channel half-section, $h_0$. Inset: zoom of the main panel.}
\label{flatwall}
\end{figure}
% 
% %              PROFILI             
\begin{figure*}[htb]
\includegraphics[scale=1.] {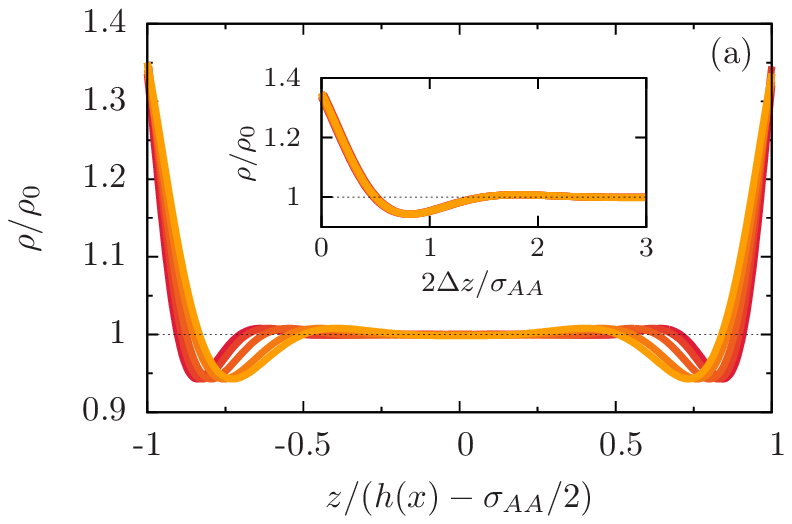}\includegraphics[scale=1.]  {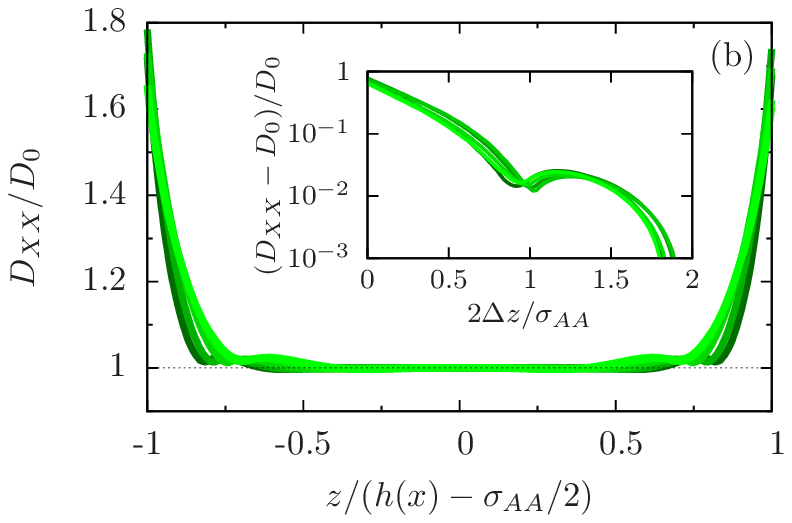}
\includegraphics[scale=1.]  {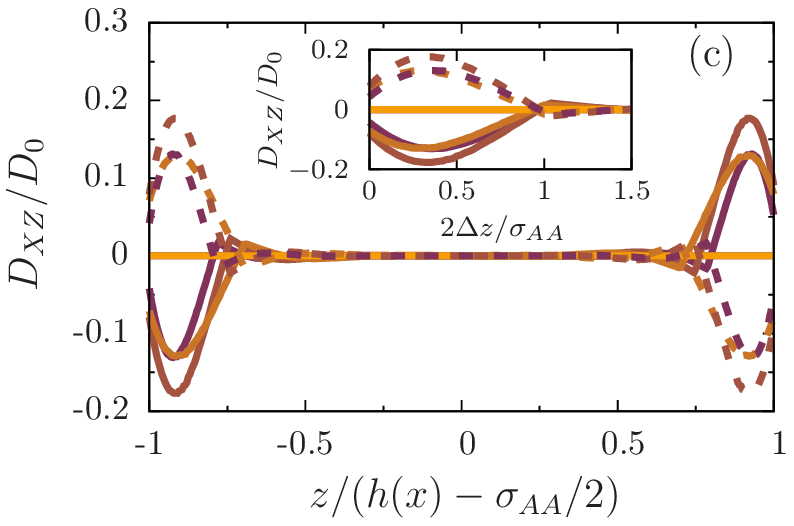}\includegraphics[scale=1.]  {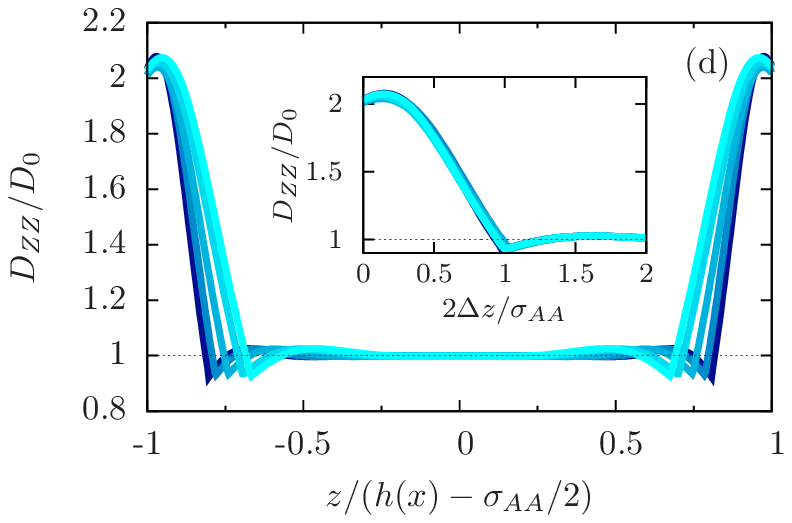}
\caption{(colors online) Corrugate walls ($\Delta S=0.2$ for all panels). a) density profile, normalized by the bulk value $\rho_0=0.2$, of $A$ particles as a function of the transverse distance, normalized by the local channel half-amplitude $h(x)$, for different longitudinal positions, $x=0,L/8,L/4,3L/8,L/2$ standing lighter colors for larger values of $x$. b), c) and d) components of the diffusion tensor profile $XX$ (panel b), $XZ$ (panel c) and $ZZ$ (panel d) component, normalized by the bulk value $D_0$, as a function of the transverse distance, normalized by the local channel half-amplitude $h(x)$, for different  longitudinal positions, $x=0,L/8,L/4,3L/8,L/2$ (solid lines) and $x=-L/8,-L/4,-3L/8$ (dashed liens) standing lighter colors for larger values of $|x|$ (Note that in panel (b) and (d)  the dashed liens coincides with the solid one for symmetry reasons). Insets: dependence of the respective diffusion tensor profile as a function of the absolute distance from the channel wall, $\Delta z=h(x)-
|z|$, normalized by the $A$ particle diameter, $\sigma_{AA}
$.}
\label{nonflatwall}
\end{figure*}
%\subsection{Technical details}
\section{Results}

After having obtained the relevant effective transport equations and density profiles for the racers along a generic channel, in this section we will analyze in more detail  the effects that  the inhomogeneities in the distribution of $A$ particles and/or the channel corrugation have on the  transport properties of the system. 
The slit is defined by the condition that the centers of the $A$ particles have their transverse coordinate $-h(x)<z<h(x)$
and for $h(x)$ we assumed the functional form
$$
h(x)=h_0+\Delta\cos\frac{2\pi}{L}x .
$$
The density profile $n^A(z,x)$ can be  obtained  independently and numerically by discretizing  Eq.~\eqref{bequil} on a two dimensional 
grid in the $(x,z)$ domain. The solution is determined by an iterative process which was terminated when the difference between the profiles relative to the n-th and (n-1)-th iteration was less than a relative tolerance $10^{-10}$.
Using the $n^A(z,x)$ profile as an input we can compute the inhomogeneous pair correlation function and the matrix elements of the
friction tensor. 

We consider first a channel with planar walls as a reference geometry where the geometrical constraint does not lead to an $x$-dependent modulation in the transport coefficient, yet the presence of a boundary affects the local density of $A$-particles, as shown in Fig.~\ref{flatwall}.b. In particular, in the vicinity of the walls, the profiles vary rapidly over molecular length scales due to the action of the confining walls. The inhomogeneous density leads to a dependence in the friction matrix. As shown in Fig.~\ref{flatwall}.(a), both elements of the friction matrix increase upon approaching walls, whereas  near the center of the slit they both approach  their bulk isotropic value given by Eq.~\eqref{gammasymmetrized1}. 
%As shown Fig.~\ref{flatwall}, the pore affects considerably the fluidity of the liquid inside a nano-channel.
By inverting the friction matrix, via Eq.~\eqref{eq:diff1} we can compute the diffusion matrix elements. As shown in  Fig.~\eqref{flatwall}.(b) the non-homogeneous density profiles lead to an increase of the local diffusion coefficients. In particular, the transverse diffusion coefficient, $D_{ZZ}$ results to be larger than the longitudinal diffusion coefficient, $D_{XX}$ as shown in Fig.~\eqref{flatwall} due to the reduced rate of collision along the normal direction caused by the presence of the wall (see Appendix B). 
$D_{XX}$ characterizes the diffusion of tracer articles along the  channel. For impermeable walls the diffusion transverse to the channel, and proportional to $D_{ZZ}$,  is essentially a transient process. 
%Due to the confinement imposed by the walls, while $D_{XX}$ is the asymptotic diffusion coefficient, $D_{ZZ}$ can be only considered as a transient diffusion coefficient since the mean square displacement along the transverse direction eventually saturates due to the confinement. 
Finally, we may expect that in the presence of wetting layers adjacent the walls the friction tensor may increase due to the presence of a region of higher density  close to the wall leading to a reduction of the local diffusion coefficient. 

We quantify the   channel corrugation through the parameter
\begin{equation}
 \Delta S=\ln\left[\dfrac{h_{max}}{h_{min}}\right].
 \label{def:DS}
\end{equation}.
%When $\Delta S\neq0$, the mesoscopic modulation of channel amplitude induces density profiles that depend on the local amplitude of the channel. 
The dependence of the density profile of $A$ particles upon channel amplitude is captured in Fig.~\ref{nonflatwall}.(a). 
As shown in the inset of Fig.~\ref{nonflatwall}.(a), for a slowly varying channel the density profiles essentially collapse in a unique curve if the  density is plotted as a function of their distance to the wall. Therefore, the  variation of channel amplitude modulates the overall density profile but not at length scales comparable to the size of $A$ particles. 

As sown in Fig.~\ref{nonflatwall}, for corrugated walls the variations in the local channel amplitude lead, on one hand to a modulation of both $D_{XX}$ and $D_{ZZ}$ (see Figs.~\ref{nonflatwall}.(b) and (d) respectively), and on the other hand to a non-vanishing contribution to the off-diagonal term $D_{XZ}$ (see Fig.~\ref{nonflatwall}.(c)).  In contrast to the density profiles, which collapse on a master curve in the vicinity of the channel wall, even though the modulations  of $D_{XX}$  and $D_{ZZ}$ approach a master curve when plotted as a function of the rescaled distance from the wall $\delta z$,  as shown in the insets of  Figs.~\ref{nonflatwall}.(b),(d), they show larger deviations from a master curve when  compared to Fig.~\ref{nonflatwall}.(a).
Analogous to the flat channel, we always find the transverse diffusion larger than the longitudinal one, $D_{ZZ}>D_{XX}$, whereas for corrugated channels the off-diagonal contribution, $D_{XZ}$, does not vanish and, as expected, has a magnitude smaller than observed for the diagonal terms of the diffusion matrix. Moreover, it is interesting to note that while $D_{XX}$ and $D_{ZZ}$ are symmetric about the channel longitudinal axis, $D_{XZ}$ is antisymmetric. 
%Such a feature is due to the channel symmetry about its longitudinal axis.
 %In fact, for increasing channel amplitudes, $B$ particles shall explore a larger channel section. 
 Therefore particles above (below) the channel longitudinal axis will experience a local ``drift'' towards the closest channel wall that reflect in a positive (negative) off-diagonal mobility $\mu=\beta D_{XZ}$ that can lead to an excess accumulation of particles on channel walls when $B$ particles are driven by an external force. 

%However, if one considers a slit where walls are impermeable to the solute by allow solute molecules to cross, then the previous expression indicates that there will be a transverse solute  diffusive flux induced by the  spatial inhomogeneity o the bounding channel.
%The magnitude of $D_{XX},D_{ZZ}$ and $D_{XZ}$ depends on the amplitude of the modulation of channel amplitude as captured by the parameter: 

% Figures ~\ref{nonflatwall}.(b),(c) and (d) show that $D_{XX}$, $D_{ZZ}$ and $D_{XZ}$ undergo larger modulations upon increase of $\delta S$. In particular, $D_{ZZ}$ is the most sensitive to $\Delta S$ and it can be as much as $50$ times larger in the channel bottleneck as compared to its bulk value while $D_{XX}$ shows a milder dependence on $\Delta S$. Interestingly $D_{XZ}$ shows a remarkable dependence on $\Delta S$.

\begin{figure*}
 \includegraphics[scale=1.]{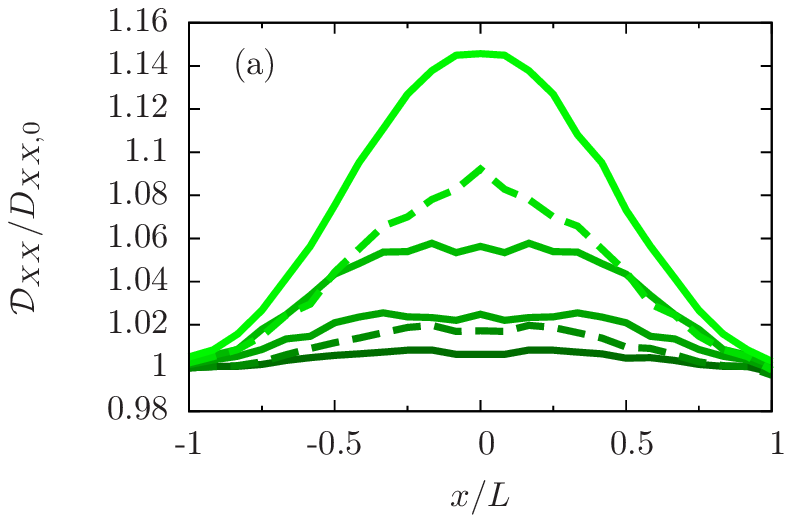}\includegraphics[scale=1.]{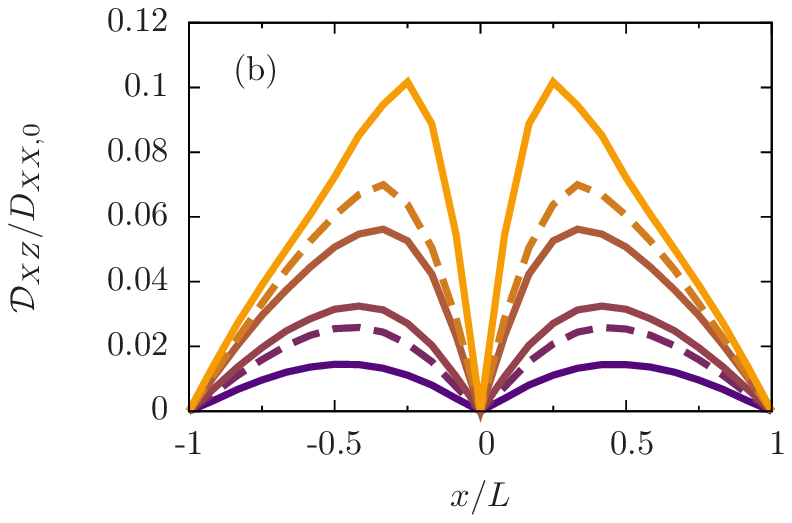}
 \includegraphics[scale=1.]{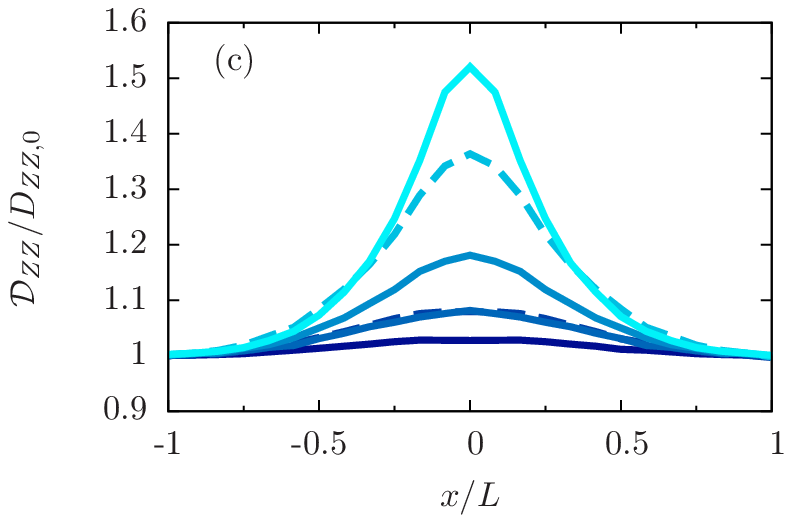}\includegraphics[scale=1.]{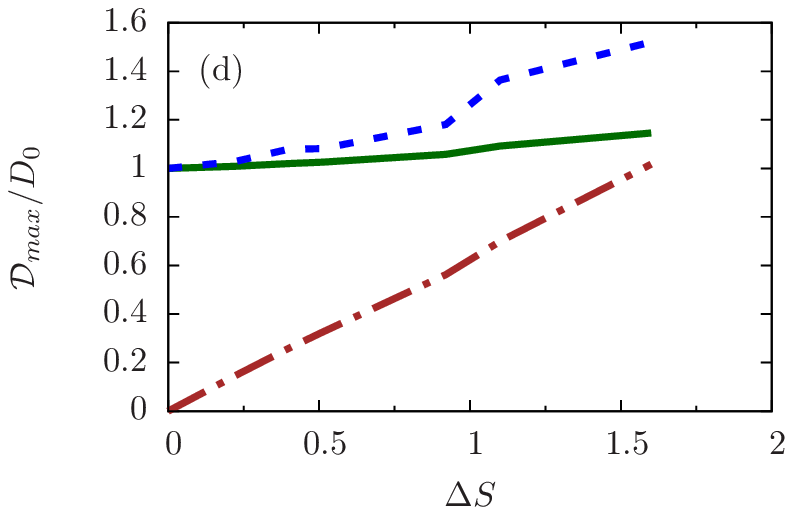}
 \caption{(colors online) Dependence of $\mathcal{D}_{XX}$ (a), $\mathcal{D}_{XZ}$ (b) and $\mathcal{D}_{ZZ}$ (c) as a function of the longitudinal position for different values of the entropic barrier $\Delta S=0.22,0.4,0.51,0.92,1.1,1.6$ standing lighter colors for larger values of $\Delta S$ (solid lines are for channels whose maximum amplitude is $10\sigma_{AA}$ while dashed liens for channel whose maximum amplitude is $6\sigma_{AA}$).  (d): dependence of the normalized maximum value of $\mathcal{D}_{XX}$ (solid line), $\mathcal{D}_{ZZ}$ (dotted line) and $\mathcal{D}_{XZ}$ (dash-dotted line, magnified by a factor of $10$ for clarity shake) as a function of $\Delta S$.}
 \label{fig:DS-dep}
\end{figure*}

Fig.~\ref{fig:DS-dep} shows that the dependence of the average diffusion coefficients, $\mathcal{D}_{XX}$,$\mathcal{D}_{ZZ}$ and $\mathcal{D}_{XZ}$, is enhanced by increasing $\Delta S$. Interestingly, both $\mathcal{D}_{XX}$ and $\mathcal{D}_{ZZ}$ attain their maximum value in the channel bottleneck, $x=0$. In particular, while $\mathcal{D}_{XX}$, Fig~\ref{fig:DS-dep}.(a), shows a broader dependence on the longitudinal coordinate, $\mathcal{D}_{ZZ}$, Fig.~\ref{fig:DS-dep}.(c) is more sensitive around the channel bottleneck, $x=0$. 
In contrast, Fig.~\ref{fig:DS-dep}.(b) shows that the maximum of $\mathcal{D}_{XZ}$ is not obtained in the channel bottleneck and that its position is $\Delta S$-dependent. 
Showing the variation of the  maximum value of the different components of the diffusion matrix as  a function of the channel corrugation highlights the impact that geometrical variations has on the diffusion of trace particles, see Fig.~\ref{fig:DS-dep}.(d). For the diagonal components  one can see increases of tens of a percent for the longitudinal component of the diffusion and up to $50\%$ for the transverse one. This plot also highlights that the off-diagonal component becomes non zero exclusively as a result of the corrugation in an interacting system and that the increase in its value is linear in $\Delta S$.

% Finally, Fig.~\ref{fig:DS-dep}.(d) shows the dependence of the maximum value of $\mathcal{D}_{XX}$,$\mathcal{D}_{ZZ}$ and $\mathcal{D}_{XZ}$ as a function of $\Delta S$. In particular we notice that. while $\mathcal{D}_{XX}$,$\mathcal{D}_{ZZ}$ have a more mild dependence on $\Delta S$, $\mathcal{D}_{XZ}$  is more sensitive to $\Delta S$ and it increases almost linearly in $\Delta S$.

Both the inhomogeneities in the density of the solvent species ($A$-particles) across the channel and the channel   corrugation  affect the net transport properties of the system. 
In particular, our framework allows us to analyze systematically the impact that the channel corrugation has on the mean particle flows. To this end, we analyze the particle flux induced by an applied, uniform external field when the channel is in contact with two reservoirs kept with  equal chemical potentials.
% In particular, our framework allows us to study different situations and, by comparison, to identify new possible scenarios induced by the inhomogeneities. 
For what concerns the dependence of the net flux on the channel geometry, imposing periodic boundary conditions, $c(-L/2)=c(L/2)$ in   Eq.~\eqref{firstorderb} and assuming the symmetry of the channel, $h(x)=h(-x)$, we obtain expressions for the flux under the action of a constant driving force, $F_0$, in the flat channel and in  the corrugated channel cases, , $I_f$ and  $I_c$ respectively:
\begin{multline}
 I_{f,c}=c_{f,c}(L/2)e^{\beta(\mathcal{A}_{f,c}(L/2)-fL/2)}(e^{\beta F_0L}-1)\cdot\\
 \cdot\Bigl[\int_{-L/2}^{L/2} \frac{e^{\beta (\mathcal{A}_{f,c}(x)-F_0x)}}{\mathcal{D}_{XX,f,c}(x)}dx \Bigl]^{-1}
 \label{flux_FJ}
\end{multline}
where $\Delta \mathcal{A}_{f,c}=\mathcal{A}_{f,c}(L)-\mathcal{A}_{f,c}(0)$ and $\mathcal{A}_{f,c}$ stands for the equilibrium free energy  for a flat or a corrugated channel, as defined in Eq.~\eqref{mu0}, and $\mathcal{D}_{XX,f,c}(x)$ equals $\mathcal{D}_{XX}$ calculated respectively for a flat or corrugated channel. 

The ratio of the mean fluxes in the corrugated and flat channels, $I_c/I_f$, gives
\begin{multline}
 \frac{I_c}{I_f}=\frac{c_c(L/2)e^{-\beta\mathcal{A}_c(L/2)}}{c_f(L/2)}e^{-\beta F_0L/2}\frac{e^{\beta F_0L}-1}{\beta \mathcal{D}_{XX,f}(L/2)f}\cdot\\
 \cdot\left[\int_{-L/2}^{L/2}\frac{e^{\beta(\mathcal{A}_c(x)-fx)}}{\mathcal{D}_{XX,c}(x)}dx\right]^{-1}
 \label{eq:flux-ratio}
\end{multline}
where we have exploited the fact that for a flat channel $\int_{-L/2}^{L/2}\frac{e^{\beta (\mathcal{A}_f(x)-F_0x)}}{\mathcal{D}_{XX,f}(x)}dx=\frac{e^{\beta \mathcal{A}_{f}(L/2)}}{\mathcal{D}_{XX,f}(L/2)}(e^{\beta F_0L}-1)$. 
For a weakly corrugated channel, $\beta\mathcal{A}\ll 1$ we can expand  the diffusion coefficients  with respect to the reference, flat channel, 
 $\mathcal{D}_{XX,c}(x)=\bar{\mathcal{D}}_{XX,c}+\tilde{\mathcal{D}}_{XX,c}(x)$. Accordingly,  we can expand the last integral  in Eq.~\eqref{eq:flux-ratio} for small $\beta(\mathcal{A}_{f}(L/2)-F_0L)$ and $\tilde{\mathcal{D}}_{XX,c}(x)$ up to second order getting:
 \begin{widetext}
\begin{equation}
 \int_{-L/2}^{L/2}\frac{e^{\beta(\mathcal{A}_c(x)-F_0x)}}{\mathcal{D}_{XX,c}(x)}dx\simeq 
 \frac{1}{\bar{\mathcal{D}}_c}\int_{-L/2}^{L/2} 1+\beta(\mathcal{A}_c(x)-F_0x)-\frac{\tilde{\mathcal{D}}_{XX,c}(x)}{\bar{\mathcal{D}}_{XX,c}}
 +\frac{1}{2}\beta^2(\mathcal{A}_c(x)-F_0x)^2+2\frac{\tilde{\mathcal{D}}^2_{XX,c}(x)}{\bar{\mathcal{D}}^2_{XX,c}}-\beta\mathcal{A}_c(x)\frac{\tilde{\mathcal{D}}_{XX,c}(x)}{\bar{\mathcal{D}}_{XX,c}}dx
 \label{eq:integr-exp1}
\end{equation}
where we have assumed  $\tilde{\mathcal{D}}_{XX,c}(x)\ll\bar{\mathcal{D}}_{XX,c}$. Without loss of generality, we can assume $\int_{-L/2}^{L/2} \tilde{\mathcal{D}}_{XX,c}(x)dx=0$. Accordingly the last expression reduces to:
\begin{equation}
 \int_{-L/2}^{L/2}\frac{e^{\beta(\mathcal{A}_c(x)-F_0x)}}{\mathcal{D}_{XX,c}(x)}dx\simeq 
 \frac{1}{\bar{\mathcal{D}}_{XX,c}}\left[L+\beta (\tilde{\mathcal{A}_c}-F_0L) 
 +\int_{-L/2}^{L/2}\frac{1}{2}\beta^2\mathcal{A}^2_c(x)+2\frac{\tilde{\mathcal{D}}^2_{XX,c}(x)}{\bar{\mathcal{D}}^2_{XX,c}}-\beta\mathcal{A}_c(x)\frac{\tilde{\mathcal{D}}_{XX,c}(x)}{\bar{\mathcal{D}}_{XX,c}}dx\right]
 \label{eq:integr-exp2}
\end{equation}
\end{widetext}
which indicates, interestingly, that the tracer flux in a corrugated channel is reduced by an amount  controlled by the modulation amplitude   in both $\mathcal{A}$ and $\mathcal{D}$. Moreover, the presence of the cross term $-\int_{-L/2}^{L/2}\beta\mathcal{A}_c(x)\frac{\tilde{\mathcal{D}}_{XX,c}(x)}{\bar{\mathcal{D}}_{XX,c}}dx$ reduces the overall flux. In principle, this contribution can lead to flux reversal, and, in any case, indicate that there will exist an optimal channel shape that maximizes the net flux of tracer particle along the channel

\section{Conclusions}

We have analyzed the impact that channel corrugation has on diffusive particle transport. To this end, we have developed  a framework that allows us to capture the coupling between the intrinsic inhomogeneities due to the microscopic properties of the medium and the  modulations in the channel amplitude. In particular, we have focused on the equilibrium as well as transport properties of a very dilute suspension (tracer limit) of particles, $B$, moving through 
an environment formed by larger particles, $A$, when the overall system is confined between two corrugated plates. 
The interplay between spatial corrugation and non-ideality leads to a new type of inhomogeneity in the effective diffusion of tracer particles along a channel, absent for ideal diffusing particles. The diffusion process is characterized by a matrix where all its components depend on the local position along the channel. Channel corrugation also leads to the development of a new, off-diagonal term in the mobility matrix for  non-ideal systems.
%As a result of the wall  corrugation, the microscopic properties of the system acquire a dependence on the longitudinal coordinate (absent in the case of flat channels) and lead to the appearance of a new, off-diagonal term in the mobility matrix. 
%and induce non-vanishing off-diagonal terms  of the mobility matrix of the $B$ particles.  

As a reference we have characterized the density profile as well as the diffusion coefficients of the solute species ($B$-particles) in a flat channel, $\Delta S=0$. Fig.~\ref{flatwall}.(b)  shows that the presence of the walls induces heterogeneities in the solvent density ($A$-particles) that ultimately affect the density and the diffusivities of $B$-particles. In particular, both the longitudinal,$D_{XX}$, and transverse, $D_{ZZ}$, diffusion coefficient increase upon approaching the wall. Such a behavior is due to the lack of $A$ particles in the vicinity of the wall that eventually decreases the effective drag experienced by the $B$ particle (see Appendix B).  

For corrugated channels, we characterize their inhomogeneity in terms of a tuning parameter that measures the difference between maximal and minimal channel  aperture, and that can be seen, effectively, as a measure of the geometric, or entropic resistance to particle flow, $\Delta S$. For corrugated channels, when $\Delta S\neq0$, the channel amplitude  modulation  leads to density profiles that depend on the channel local amplitude. Interestingly, the  variation of channel amplitude only mildly  affects  the density profile of $A$ particles. Accordingly, the density profiles for different amplitudes collapse on a master curve in a region close to channel walls. These modulations in particle $A$ density eventually determine the effective diffusion coefficients of $B$ particles. In particular, when $\Delta S\neq0$ the off-diagonal terms, $D_{XZ}$ becomes non-vanishing therefore leading to a coupling between the longitudinal and transverse transport properties inside the channel. This new transport mechanism 
may be relevant since it indicates a new transport mechanism that will  affect the effective permeability  across inhomogeneous membranes

Fig.~\ref{nonflatwall}.(c), shows the most striking effect of the mesoscopic modulation of channel amplitude, namely the appearance of  off-diagonal terms, $D_{XZ}$ in the diffusion matrix of $B$ particles. This new term appears due to a coupling between the  change in the channel cross section and the interaction between particles. It is not present for non-ideal systems when the density profiles satisfy local equilibrium. Interestingly $D_{XZ}$ is antisymmetric with respect to the transverse coordinate. Therefore, for systems under an external weak force (for which the linear response holds) we  expect an excess of accumulation of $B$ particles towards the channel walls in the first half of the channel and the opposite in the second half, compare solid and dashed lines in Fig.~\ref{nonflatwall}.(c). 

Since the  modulation of the diffusion coefficients depends on the  position of the particles along the channel, the overall relevance of the  modulation of channel amplitude on the diffusion coefficient is better captured by the average quantities, $\mathcal{D}_{XX}$, $\mathcal{D}_{XZ}$ and $\mathcal{D}_{ZZ}$. Fig.~\ref{fig:DS-dep} shows that all the components of the diffusion matrix grow upon increase of $\Delta S$. In particular, Fig.~\ref{fig:DS-dep}.(d) shows that $\mathcal{D}_{XZ}$ is the most sensitive component  to changes in the channel corrugation and that it depends linearly on $\Delta S$. 
%The magnitude of the modulation in the diffusion coefficients is shown in Fig.~\ref{fig:DS-dep}.(d). Interestingly, we have found that the magnitude of the maximum of $D_{XZ}$ depends almost linearly on $\Delta S$.
From the point of view of the impact that channel modulation has on the absolute magnitude of the diffusion coefficients for the solute, Fig.~\ref{fig:DS-dep}.(d) shows that for the system under study, namely hard sphere suspension, those modulations are of the order of some tens of a percent and, in order to increase them further more corrugated channels are needed therefore limiting the possible relevance of the phenomena describe here. In particular, here we focused on the case of hard sphere suspensions (for which an analytical form for the two point correlation function is provided) that introduce a typical length scale, namely the sphere radius $\sigma_{AA}/2$. Therefore for hard spheres suspension the bottleneck of the channel cannot be smaller than $\sigma_{AA}$, namely cannot be smaller than the characteristic length. However, recently it has been shown that when the typical length characterizing the microscopic interactions is comparable to the channel bottleneck the coupling between the 
microscopic properties of the medium and the mesoscopic properties of the channel is 
maximized~\cite{PaoloElecotrokinetics}. Therefore, we expect an amplification of the  phenomena we have described for hard sphere suspension in systems characterized by a typical interaction length that is larger than the hard-core repulsion  such as fluids, polymer suspensions or in electrolytes~\cite{PaoloElecotrokinetics} just to mention a few among others, that allows for a better overlap between the microscopic interaction length and the channel bottleneck. 

\section*{Acknowledgments}
U.M.B.M. acknowledges the support received from the
European  Science  Foundation  (ESF)  for  the  activity  entitled ``Exploring the Physics of Small Devices (EPSD)'' under
the Grant No. 3720.  I.P. acknowledges  the Direcci\'on General de Investigaci\'on (Spain) and DURSI  for financial support
under projects  FIS\ 2011-22603 and 2014SGR-922, respectively, and  {\sl Generalitat de Catalunya } under program {\sl Icrea Academia}. I.P. declares that the research leading to these results has received funding from the European Union Seventh Framework Programme (FP7-PEOPLE-2011-IIF) under grant agreement no. 301214 (\url{http://ec.europa.eu/research/fp7/index_en.cfm}).
%%%%%%%%%%%%%%%%%
%%%%%%%%%%
\appendix
\section{Boundary conditions}
Here we  show that since
in the stationary state  the B-current is solenoidal:
\be
\nabla\cdot {\bf J}^B(x,z)=0
\label{zerodiv}
\ee
the current intensity, $I_X$, along the axial direction must be uniform in space.
Integrating Eq. \eqref{zerodiv} over $z$ we obtain in the hard walls case the relation:
\be
\int_{-h(x)}^{h(x)} dz \nabla_X  J^B_X(x,z) +J^B_Z(x,h(x))-J^B_Z(x,-h(x))=0
\label{zerodivintegrated}
\ee
where the two boundary terms $J^B_Z$ vanish only in the case of flat walls whose normals are parallel to the z direction 
(condition of  impenetrability of the walls) . In order to obtain a result for non flat walls
we transform the first term of \eqref{zerodivintegrated} as:
\begin{multline}
\int_{-h(x)}^{h(x)} dz \nabla_X  J^B_X(x,z)= \nabla_X\int_{-h(x)}^{h(x)} dz J^B_X(x,z)+\\
- (J^B_X(x,h(x))+J^B_X(x,-h(x)))\nabla_X h(x)
\label{nxj}
\end{multline}
and use the vanishing of the normal flux at the walls:
\be
 {\bf J}^B(x,\pm h(x))\cdot  \hat {\bf n}(x,\pm h(x))=0
 \label{normalj}
 \ee
 where $ \hat {\bf n}$ is the normal to the wall and explicitly we have:
 \be
 J^B_X(x,\pm h)\nabla_X h(x) \mp J^B_Z(x,\pm h)=0.
 \ee
 We substitute such a  result in  Eq.\eqref{nxj}  
 and using Eq. \eqref{zerodivintegrated} we finally obtain
 that  the integral over $z$ of the $x$ component of the current ${\bf J^B}$ is 
 independent of $x$
   \begin{multline}
\int_{-h(x)}^{h(x)} dz \nabla_X  J^B_X(x,z) + J^B_Z(x,h(x))-J^B_Z(x,-h(x)) =\\
=\nabla_X\int_{-h(x)}^{h(x)} dz J^B_X(x,z)  =0
\end{multline}
that is $J_x^B$ is constant on every section
\be
\int_{-h(x)}^{h(x)} dz J^B_X(x,z)=I_X  .
\ee

\section{ Simple argument for understanding the decrease of the friction matrix elements near a  hard wall}  
  Here we give a simple argument in order to justify the decrease of the friction matrix elements
  near a planar wall. The argument is based on a rough estimate of the collision time for normal and tangential
  motions of the tagged molecules. 
 The scattering cross section for hard spheres of diameter $\sigma$ under bulk conditions is 
  $
  S=\pi \sigma^2
  $.
 If the particle moves at an average velocity $v_{th}=\sqrt{k_BT/m}$ through an assembly of hard spheres of density $n^A$
 it will suffer the following number of collisions per unit time:
 \be
 {\cal N}_{coll} = n^A   S v_{th}=\frac{1}{\tau} \ee
 where  $\tau$  is the average collision time which can be related to  the friction coefficient by
  \be
 \gamma =\frac{1}{\tau}=n^A \pi \sigma^2 v_{th} 
 \label{bulkgamma}
  \ee
  Let us remark that $\pi \sigma^2  v_{th}$ is the volume of the cylinder of influence of the particle per unit time
 and that  such a volume is reduced when the particle is at a distance from the wall less than $\sigma$.

 %The friction stems from collisions. Less collisions mean less friction.
 
% \subsection{Reduction of normal friction}
 
 It is reasonable to assume that in the proximity of a wall both the normal friction and the tangential friction 
 decrease because the tagged particle will encounter no particles in the region where the collision cylinder
 "intrudes" the wall. 
 Regarding the normal friction such an effect could be modeled by replacing 
  \eqref{bulkgamma} by
   \be
 \gamma_{Normal}(z)= \pi \sigma^2 v_{th} \frac{\int_{z-\sigma}^{z+\sigma} dz' n^A(z')}{2\sigma}
 \label{surfacegamma}
  \ee
 The rationale of such a formula is the replacement of local density $n^A(z)$ by a
 coarse grained density  obtained averaging it over a region of thickness $\sigma$, which is the region of influence of the hard sphere.
 Modeling the presence of the surface at $z=0$ by a stepwise profile,
 ($n^A(z)=0$ for $z<\sigma/2$ and $n^A_0$ otherwise) we have:
    \begin{multline}
 \gamma_{Normal}(z)= n_b \pi \sigma^2 v_{th} \frac{\int_{z-\sigma}^{z+\sigma} dz' \theta(z'-\sigma/2)}{2\sigma}
 =\\
 =n^A_0 \pi \sigma^2 v_{th} \frac{\int_{max(z-\sigma,\sigma/2)}^{z+\sigma} dz' }{2\sigma}
  \label{surfacegamma2}
  \end{multline}      
  for $z<\sigma$
    \be
 \gamma_{Normal}(z)
 =n^A_0 \pi \sigma^2 v_{th} \frac{z+\sigma/2}{2\sigma}
  \label{surfacegamma3}
  \ee        
   and for $z>\sigma$:
    \be
 \gamma_{Normal}(z)
 =n^A_0 \pi \sigma^2 v_{th} 
  \label{surfacegamma4}
  \ee          
  
 In the case of the tangential motion of the B-molecule
 the cylinder of influence will be smaller if the distance $z$ between the particle and the wall is less than $\sigma$ because 
 part of the volume of the cylinder whose axis is parallel to the x-direction is not available to the target particles.
 We know that the area $A(z)$ of the circular segment (the overlap between a circle and a half plane)
 as a function of the distance of the particle $z$ from the wall if $\sigma/2<z<\sigma$  is given by
 \be
 A(z)=\frac{1}{2} \sigma^2 (\theta-\sin\theta)  
 \ee
  where  
  \be
  \theta=2 \arccos(z/\sigma).
  \ee

  Substituting 
  \be
  A(z)=\sigma^2\Bigl( \arccos(z/\sigma)-\frac{z}{\sigma}\sqrt{1-(\frac{z}{\sigma})^2 } \Bigl)
  \label{surfacegamma-A}
  \ee 
  Thus the volume per unit time available to particle B to collide is reduced 
  \be
  V_{cyl}=\Bigl(\pi \sigma^2-A(z)\Bigl) v_{th}
  \ee
  and we find 
     \be
 \gamma_{Tangential}(z)
 =n^A_0 V_{cyl}=n_b \Bigl(\pi \sigma^2-A(z)\Bigl) v_{th}  \label{surfacegamma4-1}
  \ee    
     
%\bibliographystyle{apsrev}
%\bibliographystyle{unsrt}   % this means that the order of references
			    % is determined by the order in which the
			    % \cite and \nocite commands appear
\bibliography{umberto}  % list here all the bibliographies that
			     % you need. 
%%%%%%%%%

\end{document}